\documentclass{optica-article}
\journal{opticajournal} 

\articletype{Research Article}

\usepackage{lineno}
\usepackage{graphicx}
\usepackage{caption}
\usepackage{subcaption}

\usepackage[version=4,arrows=pgf-filled,
textfontname=sffamily,
mathfontname=mathsf]{mhchem}

\begin{document}

\title{End-to-end simulations of photonic phase correctors for adaptive optics systems}

\author{Dhwanil Patel,\authormark{1} Momen Diab,\authormark{1,*} Ross Cheriton,\authormark{2} Jacob Taylor,\authormark{1,3} Libertad Rojas,\authormark{1} Martin Vachon,\authormark{2} Dan-Xia Xu,\authormark{2} Jens H. Schmid,\authormark{2} Pavel Cheben,\authormark{2} Siegfried Janz,\authormark{2} and Suresh Sivanandam\authormark{1,3} }

\address{\authormark{1}Dunlap Institute for Astronomy and Astrophysics, University of Toronto, Toronto, Ontario, Canada\\
\authormark{2}Quantum and Nanotechnologies Research Centre, National Research Council Canada, Ottawa, Ontario, Canada\\
\authormark{3}David A. Dunlap Department of Astronomy and Astrophysics, University of Toronto, Toronto, Ontario, Canada}

\email{\authormark{*}momen.diab@utoronto.ca} 

\begin{abstract*} 
Optical beams and starlight distorted by atmospheric turbulence can be corrected with adaptive optics systems to enable efficient coupling into single-mode fibers.
Deformable mirrors, used to flatten the wavefront in astronomical telescopes, are costly, sensitive, and complex mechanical components that require careful calibration to enable high-quality imaging in astronomy, microscopy, and vision science. They are also impractical to deploy in large numbers for non-imaging applications like free-space optical communication. Here, we propose a photonic integrated circuit capable of spatially sampling the wavefront collected by the telescope and co-phasing the subapertures to maximize the flux delivered to an output single-mode fiber as the integrated photonic implementation of a deformable mirror.
We present the results of end-to-end simulations to quantify the performance of the proposed photonic solution under varying atmospheric conditions toward realizing an adaptive optics system without a deformable mirror for free-space optical receivers.\footnote{\textcopyright 2024 Optica Publishing Group. Users may use, reuse, and build upon the article, or use the article for text or data mining, so long as such uses are for non-commercial purposes and appropriate attribution is maintained. All other rights are reserved.}

\end{abstract*}

\section{Introduction}

All-optical interfacing of free-space optical (FSO) satellite-to-ground 
communication with existing fiber networks requires the coupling of the optical beams from satellite 
transmitters into single-mode fibers (SMFs). This enables high bandwidth links, optical amplification, coherent and quantum communication schemes, and long propagation distances. 
However, light propagating through Earth's turbulent atmosphere suffers from distortions that destroy its spatial coherence, prohibiting efficient coupling into SMFs. The aberrations cause the focal pattern of a point source, i.e., the point spread function (PSF), to break into an extended speckle pattern that evolves rapidly, mainly driven by the wind, which has velocities $<50~$m/s, in the case of astronomical telescopes. 
For applications in FSO communication, the temporal variations are dominated by the slewing rate of the orbiting satellite, which in the case of low Earth orbit (LEO)-to-ground links \cite{giggenbach_downlink_2022}, can reach up to $1~$deg/s, or $400~$m/s effective speeds of the optical column.
Furthermore, communication links need to be established soon after the satellite has risen above the horizon to maximize the link duration \cite{thompson_nasas_2023}, leading to highly distorted beams propagating through a larger airmass during the low-elevation stages of the link.

Adaptive optics (AO) systems in astronomical telescopes use wavefront sensors (WFSs) to sample the wavefront and feed commands, through a controller, to deformable mirrors (DMs) that change shape to correct the distorted wavefronts.  
DMs are mechanical in nature and tend to be too costly for many applications outside of astronomy, like the large-scale deployment of optical ground stations (OGSs) to serve rural and remote communities.
DMs are also mechanically limited in speed and stroke, inhibiting their use in situations that require higher correction bandwidths, like LEO-to-ground FSO links. Replacing bulk optics with integrated optics and photonic components has regularly been an alternative for overcoming such challenges in optical systems.

Photonic technologies for processing light in waveguides and optical fibers have been developed for several applications, mainly driven by the demands of the telecommunication industry in fiber-to-copper transceivers.
The photonic approach offers a platform to realize devices that can overcome the limitations of bulk optics instruments, producing solutions that are compact, scalable, lower in cost, and easier to replicate. Photonic spectrographs \cite{stoll_high-resolution_2017}, integrated beam combiners \cite{minardi_discrete_2019}, fiber-based hydroxyl-suppression filters \cite{ellis_gnosis:_2010}, and many other concepts have been developed for astronomical applications \cite{jovanovic_2023_2023}. However, the poor coupling of starlight and the low total throughput of these astrophotonic devices remains a challenge that needs to be addressed before they are adopted as facility instruments.   
In AO systems, several concepts for photonic wavefront sensors (WFSs) have been suggested as focal plane sensors that can detect non-common path aberrations (NCPAs) and petal modes in astronomical telescopes \cite{diab_modal_2019, norris_all-photonic_2020}. 
Wavefront correctors based on multiplane light converters (MPLC) have also been developed for FSO links, which use phase plates to sequentially decompose the light into orthogonal modes \cite{billaud_10_2023}, and integrated circuits for coherent combination \cite{billault_free_2021}. However, the strategy still involves bulk optic phase plates for spatial demultiplexing.

\begin{figure}[hbt!]
    \centering
    \includegraphics[width = \textwidth]{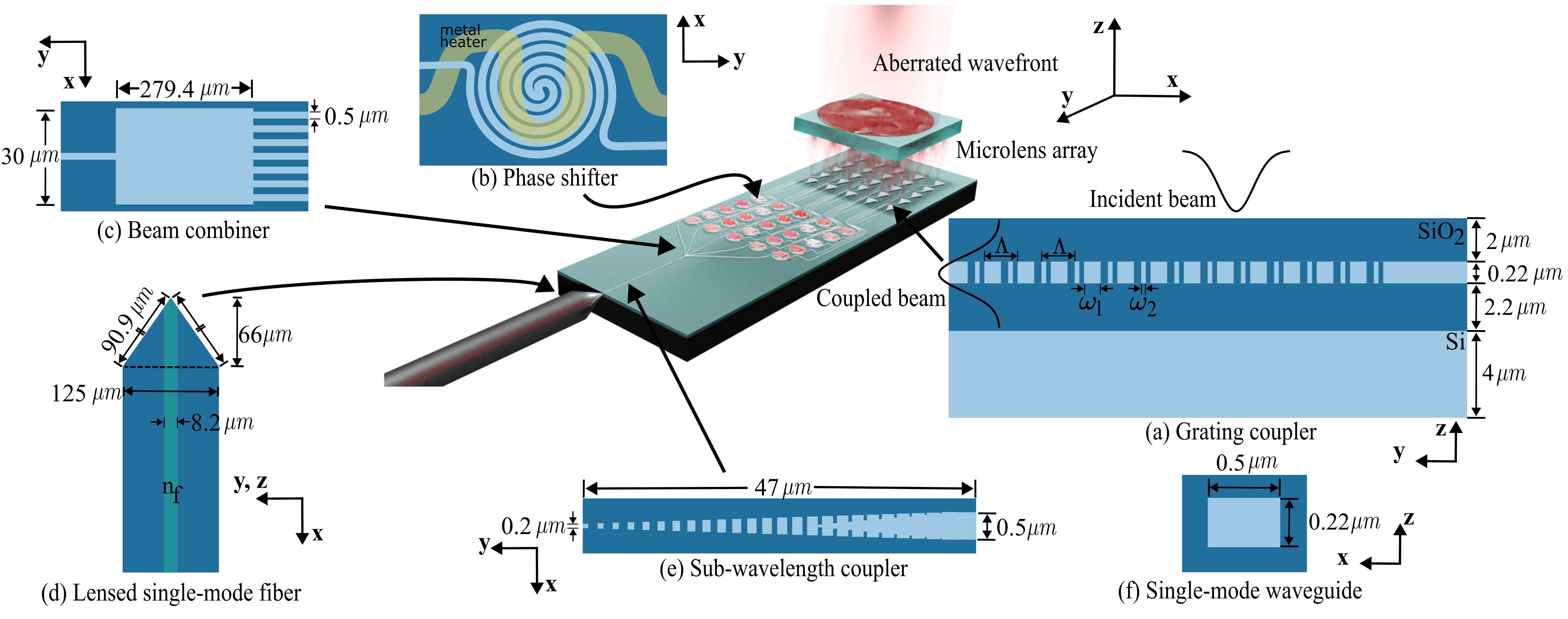}
    \caption{Schematic of the integrated chip, with a microlens array to sample and focus the distorted wavefront on the grating coupler array (a). Co-phased beams from the TOPSs (b) are coherently combined using an MMI beam combiner like the $7 \times 1$ combiner shown in (c). The combined beams are coupled out to a lensed SMF (d) using an SWG coupler (e) that matches the waveguide mode to the fiber mode. The cross-section of the single-mode waveguides is shown in (f).}
    \label{fig: pic_schematic}
\end{figure}

In this work, we propose a photonic integrated circuit (PIC) that spatially samples the light from the exit pupil of an optical ground station (OGS) and corrects the phase distortions in the incoming wavefront to boost the coupling efficiency into SMFs before amplification, transmission, demodulation, and eventual detection. 
The silicon-on-insulator (SOI) chip, shown in Fig. \ref{fig: pic_schematic}, incorporates an array of vertical incidence grating couplers that direct light focused by a microlens array (MLA) into the plane of the PIC single-mode waveguides.
Light from each spatial channel is phase-shifted using thermo-optic microheaters to modulate the refractive index and thus change the optical path length (OPL).
An external controller, fed by a wavefront sensor, drives the heaters to match the modes in phase. 
The co-phased beamlets are combined into one single-mode waveguide using a multimode interferometer (MMI) beam combiner. Finally, a sub-wavelength grating (SWG)\cite{cheben_broadband_2015} expands the propagating mode and couples the light out to an SMF. 
Using thermo-optic phase shifters (TOPSs) in combination with spiral \cite{densmore_compact_2009} or serpentine waveguides provides up to $100~$kHz control bandwidth and a phase shift $>10~\mu$m, 
exceeding what is possible with DMs based on voice coils or micro-electromechanical systems (MEMS).
The size of the array, i.e., the number of controlled subapertures, also scales more gradually with cost in the case of the photonic corrector compared to DMs.
Other advantages include a smaller footprint, lower power consumption, and a multiplexing advantage that could be applied in fiber-fed astrophotonic instruments \cite{diab_starlight_2021, jovanovic_2023_2023} and multi-object spectroscopy (MOS) applications in astronomy \cite{diab_photonic_2023}.

We previously presented proof-of-concept simulations of a generic system with square arrays and assumed idealized models for the coupling and combination components in \cite{diab2022photonic}. 
In this work, we report on the results of end-to-end simulations for the expected performance of the proposed photonic solution.
The models for the grating couplers and the waveguide-to-fiber couplers match the design of a fabricated chip that we plan to use in lab and field experiments.
We also investigate various turbulence and link scenarios, accounting for scintillation and fill factor effects on a simulated system. In Sec. \ref{sec: simpipe}, we describe the numerical tools and methods used in the simulation pipeline. 
Section \ref{sec: pd} details the geometry of the circuit components and presents a characterization of the tolerances and the spectral range of the elements. 
The simulation results are given in Sec. \ref{sec: end2eresults} for devices of different geometries and various turbulence conditions. 
A discussion of the results and their significance to the design process is also provided in that section.

\section{Optical Simulation Pipeline}
\label{sec: simpipe}
\begin{figure}[hbt!]
    \centering
    \includegraphics[width = \textwidth]{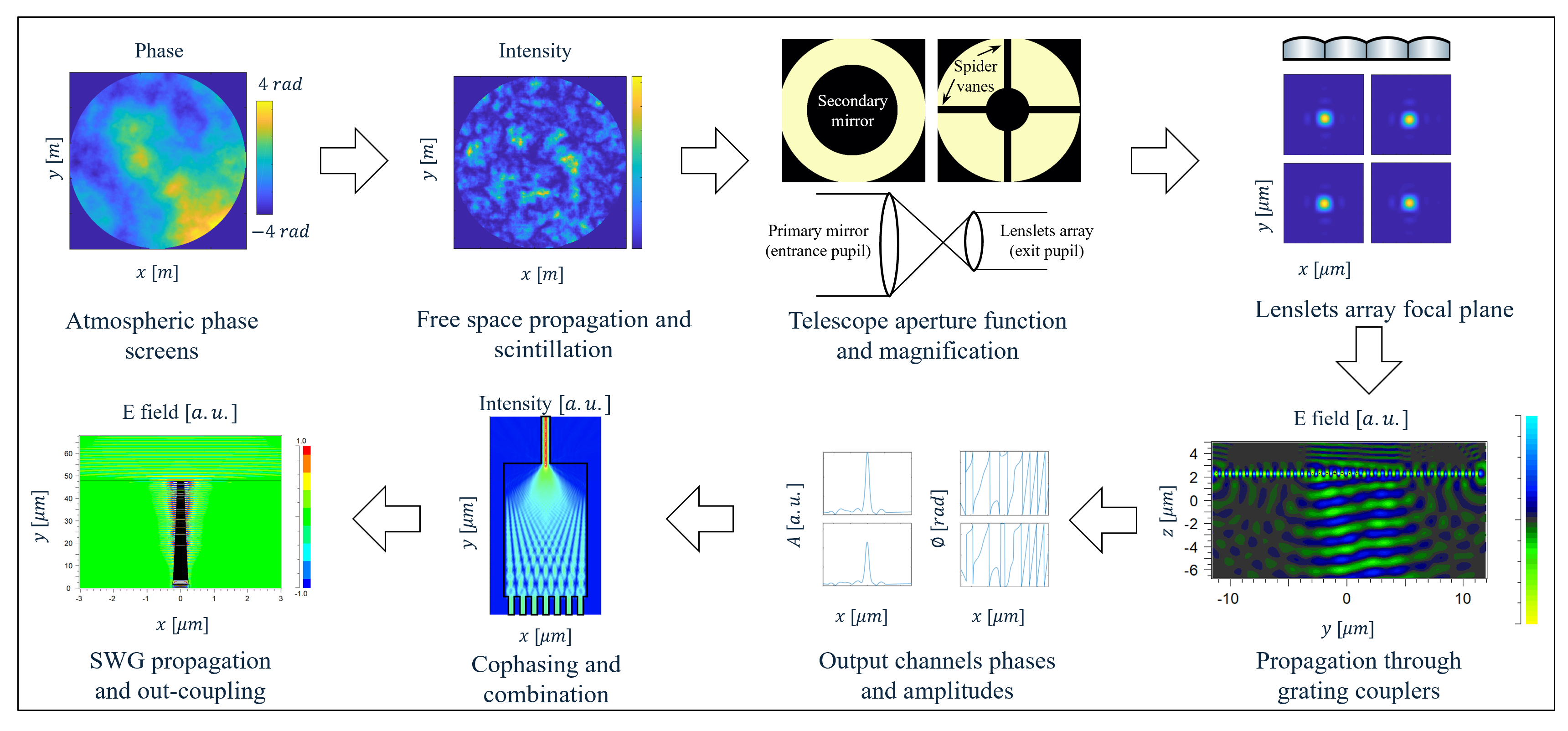}
    \caption{Optical simulations pipeline. A generator computes distorted phase screens with Kolmogorov statistics, followed by a wave optics propagator to calculate the fields at the telescope pupil. The spots at the focal plane of a lenslet array where the PIC is aligned are then computed. An FDTD calculator propagates the focal fields into a 2D model of the grating couplers. The output fields are co-phased and then combined by the MMI. The coupling efficiency into the output SMF is calculated by propagating the field through the SWG structure to the chip’s facet and into the lensed SMF.}
    \label{fig: end-to-end simulation pipeline}
\end{figure}

The simulation results we present in this work quantify the performance of the photonic wavefront corrector assuming an ideal WFS and controller. The results give the expected optical power collected at the output SMF for the scenarios encountered in a LEO-to-ground FSO link. To do so, the pipeline begins by propagating the optical wavefronts from the top of the atmosphere down to the output SMF.
The simulation pipeline in Fig. \ref{fig: end-to-end simulation pipeline} includes both the free space and guided wave parts of the system, and proceeds as follows:

\begin{enumerate}
  \item Atmospheric phase generation, free space propagation, and scintillation
  \item Telescope and relay optics
  \item Microlens array focal plane
  \item Surface grating coupler
  \item Optical beam phasing and beam combiner simulation
  \item SWG mode converter and edge coupler
\end{enumerate}

\subsection{Atmospheric phase generation, free space propagation, and scintillation}

\begin{figure}[hbt!]
    \centering
        \includegraphics[width=\textwidth]{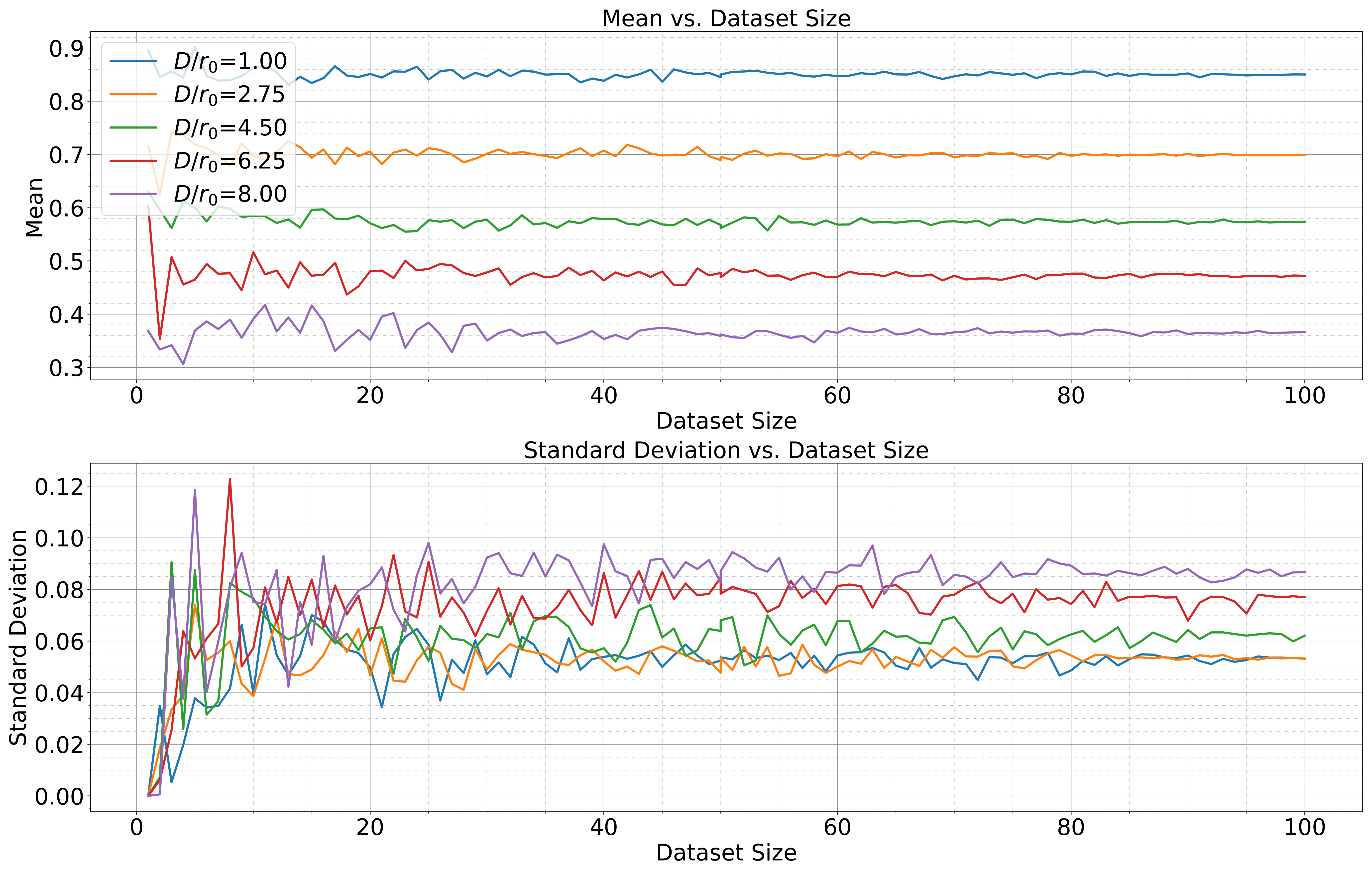}
    \caption{The mean and the standard deviation of the performance metric ($\mathit{SR}_\mathit{ph}$, described in Sec. \ref{sec: end2eresults}) for an increasing number of randomly-sampled phase screens for devices with 61 hexagons arrangement and in the presence of scintillation.}
    \label{fig: resample}
\end{figure}

As shown in Fig. \ref{fig: end-to-end simulation pipeline}, The simulation pipeline begins by generating an ensemble of phase screens representing wavefronts distorted by atmospheric layers at a given altitude. 
The screens have a von K\'arm\'an phase power spectral density \cite{welsh1997fourier}

\begin{equation}
    \Phi_{\phi}(f_s) = 0.023 r_0^{-5/3} \left( f_s^2 + \frac{1}{L_0^2} \right)^{-11/6},
\end{equation}
where the Fried parameter $r_0$ is the atmosphere's coherence length, $f_s$ is the spatial frequency and $L_0$ is the outer scale which is the maximum vortices size at which energy is supplied to the atmosphere. We take $L_0 = 22~$m \cite{maire_measurements_2007} for the results in Sec. \ref{sec: end2eresults}.
The metrics for the correction quality and the coupling efficiency of the PIC are calculated for a wide range of turbulence strength values to cover all the scenarios encountered in a LEO-to-ground link as the satellite passes from its lowest to highest elevations. 
When the link is first established, the Fried parameter $r_0$ is the smallest, while at zenith the turbulence is weaker and the satellite slew rate is at its maximum. 
We assume Taylor's frozen flow hypothesis \cite{taylor_spectrum_1938}, which states that the temporal evolution of the phase distortions is driven by the transverse effective velocities of the wind and the source rather than the inherent dynamics from temperature fluctuations. 
The phase screens are propagated through free space along the line of sight to the aperture using the Fresnel diffraction integral. 
The generated phase screens are $3$ times larger than the telescope aperture to reduce aliasing in Fresnel propagation. 
For a LEO satellite with a $422~$km altitude (ISS-like orbit), the maximum link distance is $\sim 1500~$km at a $10~$deg elevation angle while the distance to the tropopause, the highest turbulence layer, is $\sim 50~$km. 
The long propagation distance between the highest atmospheric layer and the OGS causes intensity fluctuations at the pupil, a scintillation effect that further complicates the wavefront sensing and correction process. 
Given the pseudorandom nature of atmospheric turbulence, the numerical results in Sec. \ref{sec: end2eresults} are averages obtained from a Monte Carlo simulation that calculates the performance metrics for a large sample of random realizations of the atmospheric layer.
The sample size of the Monte Carlo phase screens is $100$ realizations, chosen since it is the sample size at which the standard deviations in our performance metrics converge to their true values as shown in Fig. \ref{fig: resample}.

\subsection{Telescope optics, and relay optics}
The diameter of the collecting telescope is assumed to be $D = 40~$cm, a representative aperture size for OGSs. 
In our concept, the MLA feeding the PIC samples the pupil spatially which means that the channels that are to be co-phased and combined vary in intensity due to scintillation, degrading the efficiency of the coherent combination scheme. 
Therefore, we include the scintillation effect in the simulation pipeline to assess how it limits the device's performance. 

Subsequently, the aperture function that defines the size and arrangement of the secondary mirror and its spiders is applied to the propagated fields. 
However, the results given in Sec. \ref{sec: end2eresults} are for a clear aperture to maintain their generality.
A ray trace of the relay optics is also performed to assess the effect of the aberrations of off-the-shelf optics on the quality of the focal spots at the grating couplers. 
The resulting complex field is demagnified and imaged at the exit pupil of the optical system where the MLA is ideally placed. 

\subsection{Microlens array focal plane}
\label{subsec:MLA}
The MLA is used to spatially sample and focus subapertures from the telescope's exit pupil onto the 2D array of grating couplers in the PIC. 
Coupling with an MLA instead of directly intercepting the beam with the grating coupler array at the demagnified pupil adds an alignment step, but greatly enhances the system throughput. 
Furthermore, the configuration of the grating coupler array must match that of the MLA in size and format.

We take the MLA to have a $300~\mu$m pitch and a $1~$mm focal length, which matches the specifications of commercially available arrays. Ideally, the pitch of the MLA should be $\leq r_0$ on the sky.
We also consider two configurations for the MLA and the corresponding array of grating couplers, a square and a hexagonal arrangement of the subapertures. 
While the hexagonal configuration has a better fill factor on the circular aperture, the routing of the waveguides in the PIC is more straightforward for the orthogonal square array, especially as the size of the array grows larger.
A $25\%$ threshold is applied after masking with the aperture function to determine the partially illuminated subapertures that are not receiving enough optical power and therefore neglect them since their inclusion adversely affects the efficiency of the combination.   


\section{Photonic components design and simulations}
\label{sec: pd}  

The models for the photonic components are described here, together with simulation results of the spectral response of each component and their tolerance to fabrication inconsistencies.
As shown in Fig. \ref{fig: pic_schematic}, the simulated PIC consists of four main components. 
The grating couplers inject the spots focused by the MLA into single-mode waveguides in the plane of the PIC. 
Afterward, the TOPSs and the MMI co-phase and combine the modes into one single-mode waveguide, respectively. 
The last component is the SWG mode expander that enlarges the mode in diameter to $3~\mu$m  to efficiently couple the light out into a lensed SMF \cite{cheben_broadband_2015}.
We used a finite-difference time-domain (FDTD) solver and the beam propagation method (BPM) to model the photonic components. 
An operating central wavelength $\lambda = 1550~$nm is assumed for the optimization process since it is at the overlap of the wavelength range of SOI components, the astronomical H band which is an atmospheric transmission window, and the C band of fiber optic and FSO communication links \cite{thompson_nasas_2023}.

The PIC is based on the $220~$nm SOI photonics platform, with $500~$nm wide Si waveguides.
The buried oxide  (BOX) layer is $2~\mu$m thick SiO$_2$ 
and the top oxide (TOX) cladding is $2.2~\mu$m thick. 
These specifications are in line with the SOI process of typical Si photonic foundry offerings.  
High-resistance metal layers, e.g., TiW, are assumed for the heaters and the electrical contacts, but they are not incorporated in the optical simulations in this work. 

The complex fields at the MLA that pass the threshold test described above are propagated to the focal plane by performing a Fraunhofer diffraction integral. 
The 2D focal fields are line-scanned and the 1D profiles are used as launch fields in the 2D-FDTD solver to propagate the beamlets through to the grating couplers model in the array. 
Using a 2D-FDTD solver reduces the computation time significantly with negligible impact on the accuracy of the results \cite{yang_high-performance_2023}.


\subsection{Surface grating couplers}

Grating couplers are a modulation in the refractive index that is etched onto the waveguide to couple the free-space wave to the guided wave \cite{Katzir_chirped_1977}. 
They are typically designed to interface PICs with tilted fibers, however, our concept requires the coupling of normally incident beams from free space. 
Therefore, the grating design adopted here couples vertically incident transverse electric (TE) polarized light into the waveguides \cite{janz_optical_2023}.

The grating couplers on the PIC consist of fully etched double gratings with equal periods $\Lambda = 648~$nm and widths $\omega_1 = 363~$nm and $\omega_2 = 86~$nm, giving duty cycles of $\sim 56 \%$ and $\sim 13 \%$, respectively. Figure \ref{fig: pic_schematic}a shows a schematic of the grating coupler model. 

\begin{figure}[hbt!]
     \centering
     \begin{subfigure}[t]{0.49\textwidth}
         \centering
        \includegraphics[width=\linewidth]{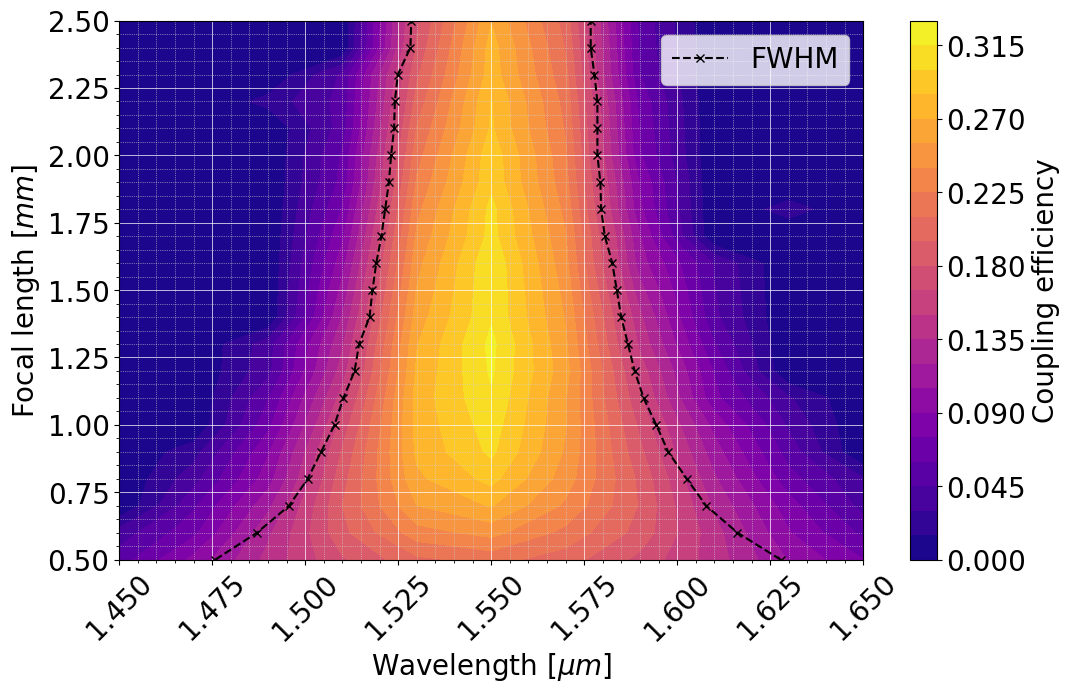}
        \caption{}
        \label{subfig: CEvsFL}
     \end{subfigure}
     \hfill
     \begin{subfigure}[t]{0.49\textwidth}
        \centering
        \includegraphics[width=0.8\linewidth]{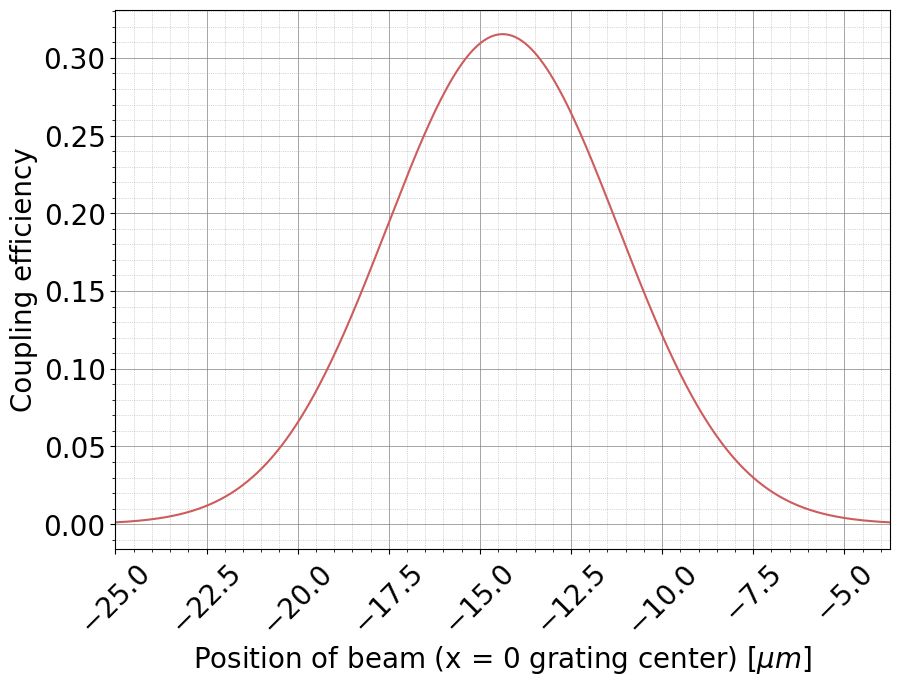}
        \caption{}
        \label{subfig: launch_pos}
     \end{subfigure}
     \hfill
     \begin{subfigure}[t]{0.49\textwidth}
        \centering
        \includegraphics[width=0.85\linewidth]{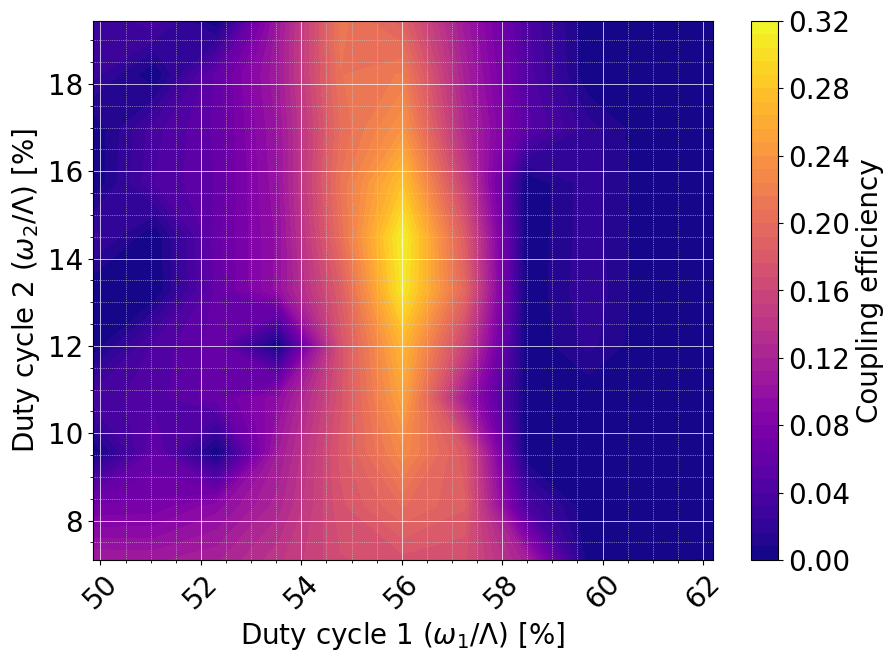}
        \caption{}
        \label{subfig: fillfactor}
     \end{subfigure}
\hfill
     \begin{subfigure}[t]{0.49\textwidth}

     \centering
        \includegraphics[width=0.9\linewidth]{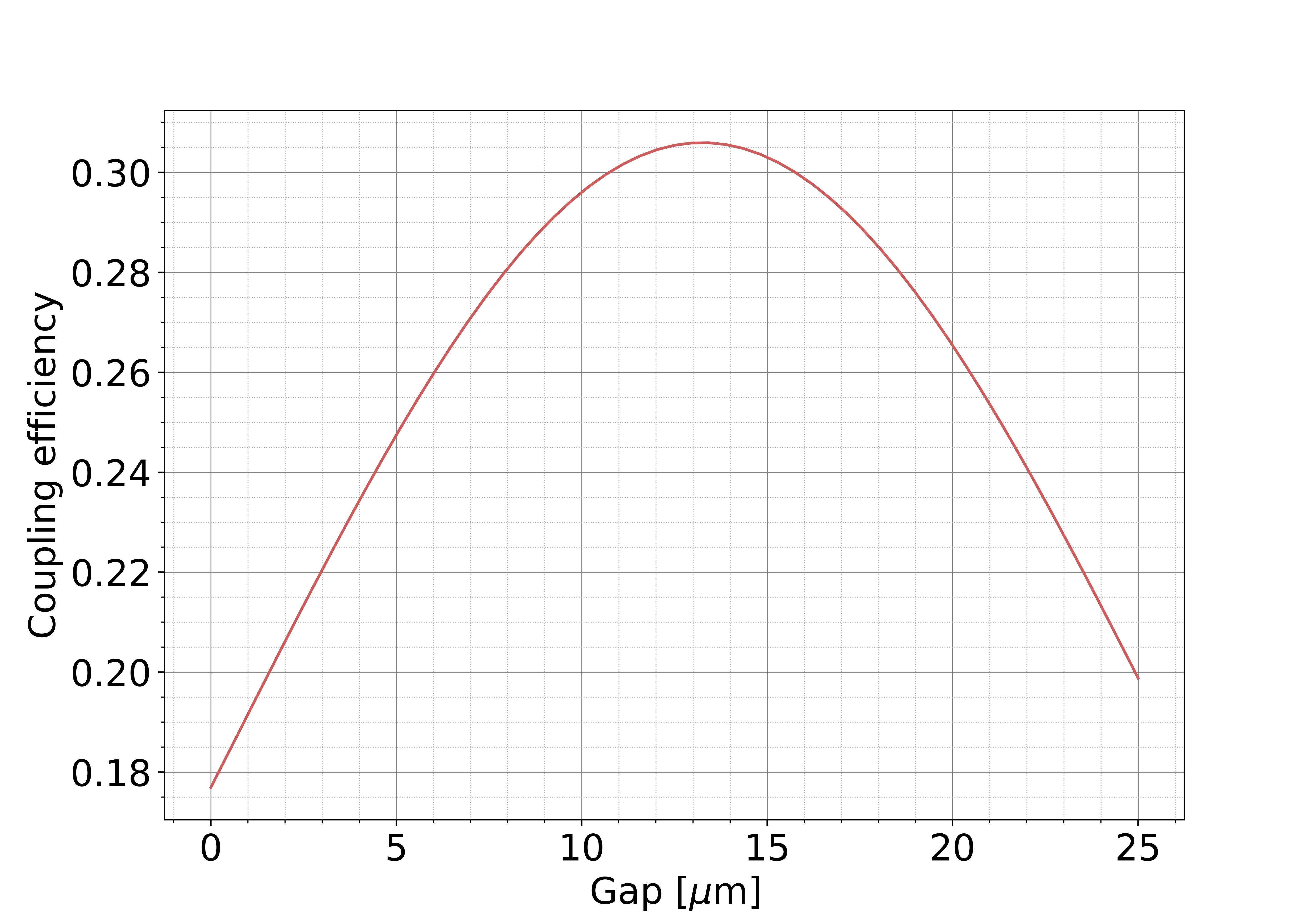}
        \caption{}
        \label{subfig: SWG_SMF}
        \end{subfigure}
    \caption{Characterization and tolerances of the grating couplers and the SWG mode expander. (a) Spectral range of the grating couplers as a function of the MLA focal length. The isolines trace the FWHM points and indicate the spectral bandwidth. The design wavelength is $\lambda = 1550~$nm and the maximum efficiency is obtained at $f= 1.25~$mm. (b) Dependence of the coupling efficiency of the grating couplers on the launch position of the focal spot. The center of the grating structure is at $x=0$ and coupling is in the negative $x$-direction. (c) Tolerance of the grating coupler to the duty cycles of the two interleaved gratings. The widths $\omega_1$ and $\omega_2$, and the period $\Lambda$ are indicated in Fig. \ref{fig: pic_schematic}a. (d) Coupling efficiency as a function of the gap between the SWG and the lensed single-mode fiber. Maximum outcoupling of $30.05\%$ is achieved at a working distance $14~\mu$m.}
    \label{fig:gc_design_tolerance}
\end{figure}

The grating coupler is optimized for coupling a vertically incident converging beam focused by an MLA with a focal length of $1~$mm at $\lambda = 1550~$nm, which gives a spot size of $\sim 12.6~\mu$m. 
The spectral bandwidth dependence on the focal length is shown in Fig. \ref{fig:gc_design_tolerance}a. 
The gratings extend to $35~\mu$m in length with the maximum efficiency obtained when the focal spot is incident $\sim 14~\mu$m off-center toward the waveguide as shown in Fig. \ref{fig:gc_design_tolerance}b.
The performance of the surface grating couplers is calculated using a 2D FDTD solver with a 2D model of the structure described. 
The duty cycles of the two gratings are scanned to estimate the penalty in coupling efficiency that we get as fabrication deviates from the design.
Figure \ref{subfig: fillfactor} shows the coupling efficiency tolerance to the changes in the duty cycles of the smaller and the bigger gratings in the $x-$ and $y-$axes, respectively. 


\subsection{Beam combiner}

The beam combiner is an MMI that combines the co-phased waveguides in its multimode region through the self-imaging principle. 
The interference between the eigenmodes of the wide waveguide produces a pattern that varies along the MMI, allowing for a design that couples all the light into one output single-mode waveguide \cite{soldano_optical_1995}. 
The design can only be optimized for one state of phase and amplitude distributions of the input waveguides. 
Therefore, optical losses would occur whenever the MMI is fed with any other state. 

\begin{figure}[hbt!]
    \centering
    \includegraphics[width = 0.75\textwidth]{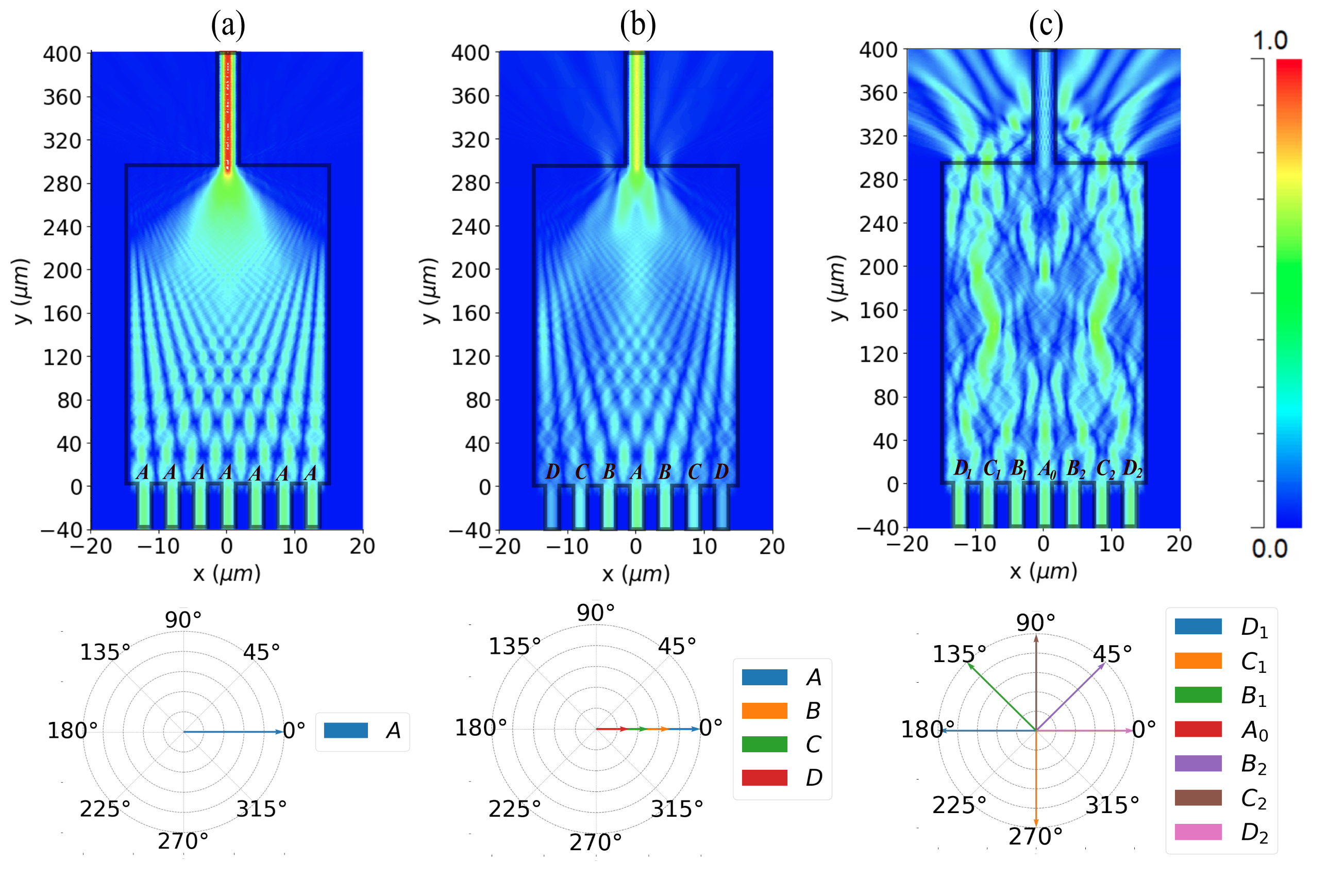}
    \caption{Intensity maps of the beams propagating through the $7\times1$ MMI combiner, with the respective phasor diagrams for the input fields. The phases of the input beamlets are indicated relative to the reference, i.e., the in-phase condition, that produces the maximum combination efficiency. (a) The beamlets are co-phased and matched in amplitude to give a combination efficiency of $99.3 \%$. (b) The beamlets are co-phased but mismatched in amplitude giving a combination efficiency of $96.6\%$. (c)  The out-of-phase condition results in very poor combining efficiency.}
    \label{fig: beamcombiner}
\end{figure}

The efficiency of the beam combiner is calculated using BPM. 
An $N\times 1$ MMI combines $N$ single-mode waveguides into one output beam. 
The design of the MMI combiners follows \cite{hosseini2009output} and an example of the field propagation through a $7\times 1$ combiner is shown in Fig. \ref{fig: beamcombiner}. 
Since the amplitudes of the modes to be combined are not always uniform due to scintillation and fill factor effects as discussed later, the impact these effects have on the combiner efficiency is also investigated.
Figure \ref{fig: beamcombiner} shows the light intensity along the beam combiner for uniform and non-uniform amplitude and phase distributions.  The maximum combination efficiency ($99.3\%$) is obtained for matched amplitudes and the following phases for the $7$ inputs: $230.7$ deg, $102.3$ deg, $25.5$ deg, $0$ deg, $25.5$ deg, $102.3$ deg, and $230.7$ deg, starting from the leftmost waveguide \cite{hosseini2009output}. As seen in Fig. \ref{fig: beamcombiner}a and \ref{fig: beamcombiner}b, the combiner exhibits high efficiencies for co-phased beams with a weak dependence on amplitude distribution. 
For the large arrays, e.g., $8\times 8$ subapertures, the simulation of the $64\times1$ MMI is computationally expensive. We therefore combine the beamlets for the larger arrays in multiple stages in a binary tree of smaller MMIs, e.g., two stages of $8\times1$ MMIs. Both combination schemes, i.e., all-in-one MMIs and MMI binary trees can be realized on SOI PICs.

\subsection{Edge coupler}

The core size of the Si waveguide in the PIC 
is an order of magnitude smaller than the mode of the output lensed SMF. 
A taper structure is therefore needed to expand the beam before interfacing with the fiber. 
A subwavelength-scale grating with a period smaller than the Bragg period expands the waveguide mode to match the fiber mode. 
The width of the SWG structure tapers down gradually towards the facet and the duty cycle smoothly decreases, reducing the effective refractive index and therefore expanding the mode \cite{cheben_subwavelength_2006}.    

The FDTD model of the SWG is shown in Fig. \ref{fig: pic_schematic}e.
It adiabatically expands the mode diameter of the $\sim500~$nm wide single-mode waveguide in the PIC to the lensed SMF aligned at the output facet \cite{cheben_refractive_2010}. 
The SWG model is required to estimate the coupling efficiency of the PIC-to-SMF segment and investigate its dependence on wavelength and air gap misalignment. 
Figure \ref{subfig: SWG_SMF} shows the dependence of the SWG to the SMF power transfer on the air gap assuming an off-the-shelf lensed fiber with a mode field diameter (MFD) of $2.5~\mu$m and a working distance of $14~\mu$m. 
The air gap is necessary to allow the beam to evolve in free-space from its size at the facet to match the mode size of the lensed fiber at the working distance. 


The total output intensity calculated from the output field coupled into the SMF is used in the next section to quantify the total efficiency of the photonic wavefront corrector. 

\section{Performance metrics results and discussion}
\label{sec: end2eresults}

The performance of any AO system is limited by instrumental errors contributed by the three main components of the system: the corrector, the WFS, and the controller. 
The WFS introduces errors resulting from the finite number of measurements it takes, the noise-limited detector, the angular separation between the reference and the target (i.e., anisoplanatism), and the cone effect in the case of laser guide stars. 
Only the first two are relevant to FSO links.
The controller's limited bandwidth causes a delay between the time the measurements were acquired and the time the correction is applied, thus introducing temporal errors. 
The wavefront corrector also has a non-zero response time that adds to the temporal errors, but most importantly it adds a fitting error since its ability to match, and thus flatten, the distorted wavefront is limited by the finite number of degrees of freedom it has.
Our interest here is in the ability of the PIC to correct atmospheric distortions and the losses inherent to SOI components.
While TOPSs always have a smaller bandwidth than electro-optic modulators, the bandwidth required in LEO laser downlinks is $< 10~$kHz during the worst-case scenario stages of the link. 
Therefore the photonic corrector is only limited by its throughput and correction degrees of freedom.

The dependence of the coupling efficiency on focal length (see Fig. \ref{subfig: CEvsFL}) is a direct result of the focal spot increasing in size at the coupling plane with longer focal lengths.
The weak dependence relaxes the design requirements of the grating couplers and means that off-the-shelf MLAs could be conveniently used. 
Figure \ref{subfig: CEvsFL} also shows the opportunity at shorter focal lengths for trading off maximum efficiency at the central wavelength for a wider spectral bandwidth. The short focal length also means that lateral motions of the spots across the grating couplers, due to tip/tilt errors in the incident wavefront, are minimal, ensuring that the focal spots always remain within the active coupling area of the gratings [see Patel et al. (submitted)].   
The tolerance of the grating coupler's efficiency to offsets in the duty cycles of the interleaved gratings is shown in Fig. \ref{subfig: fillfactor}. 
The tolerance to the relative position of the focused beam to the grating is more relaxed with a $\sim 4\%$ penalty in efficiency for a $1~\mu$m offset as shown in Fig. \ref{subfig: launch_pos}.

The relatively low $32\%$ coupling efficiency is a consequence of restricting the grating couplers' design to the common fabrication processes offered by most silicon photonics foundries. 
Adding Si overlays to the grating notches \cite{vermeulen_high-efficiency_2010}, switching to low-index core materials (e.g., Si$_3$N$_4$) apodizing the grating in the coupling direction \cite{marchetti_high-efficiency_2017}, and including Bragg reflectors \cite{mekis_scaling_2012} are all strategies that can help improve the free space-to-chip coupling efficiency up to $80\%$. 

The other point of significant loss occurs at the off-chip coupling to the SMF. 
The narrow size of single-mode Si waveguides in SOI chips results in the mode size being significantly smaller than the $\sim 10~\mu$m MFD of a silica SMF at $1550~$nm. 
Furthermore, the thin BOX layer prevents the use of direct tapers to expand the mode in the vertical dimension to match the fiber since the mode would overlap with the Si substrate. Therefore, the SWG mode expander is used to couple the light efficiently out of the PIC.
The fiber used for coupling is aligned at a working distance from the output facet, allowing the beam to diverge in free space to the correct size at the coupling plane. 
The spectral range of the SWG coupler is $>100~$nm \cite{cheben_broadband_2015} and Fig. \ref{subfig: SWG_SMF} shows the tolerance of the coupling efficiency to the air gap between the output facet and the lensed fiber. 
The relatively low efficiency is a result of assuming a sub-optimal off-the-shelf lensed SMF for the simulations, in the interest of estimating the performance of future lab experiments that use them. 
However, either an optimized SWG or a matching SMF would produce a coupling efficiency better than $92\%$. Interfacing the fiber to the PIC with a grating coupler is possible but only with a coupling efficiency $<80\%$ and with a complex grating. 

The simulation pipeline in Fig. \ref{fig: end-to-end simulation pipeline} includes all the components the beam propagates through in our device except the phase shifters. 
They are omitted since they are active components that can be precisely controlled, limited only by the resolution of the digital-to-analog conversion and the noise in the driving electronics.
The propagation and bending losses in the PIC depend on the exact routing of the waveguides in the circuit and can be easily accumulated to estimate the total throughput for a given final design.

As a performance metric, we borrow the concept of the Strehl ratio (SR) used to quantify the quality of seeing-limited PSFs in astronomy. 
We define $\mathit{SR}_{\mathit{ph}}$ (where the subscript $\mathit{ph}$ is short for \textit{photonic}) as the total optical power in the output SMF under seeing-limited conditions relative to that at the diffraction-limited case.  
Strictly speaking, our photonic wavefront corrector does not flatten the wavefront propagating in free space or restore the quality of the image at the focal plane as is the case in imaging systems.
Nevertheless, since the coupling efficiency in an SMF is directly related to the $\mathit{SR}$ \cite{faucherre_using_2000, diab_starlight_2021}, the same metric can be used for both AO concepts. 

\begin{figure}[hbt!]
    \centering
     \begin{subfigure}[b]{0.49\textwidth}
        \centering
        \includegraphics[width=\linewidth]{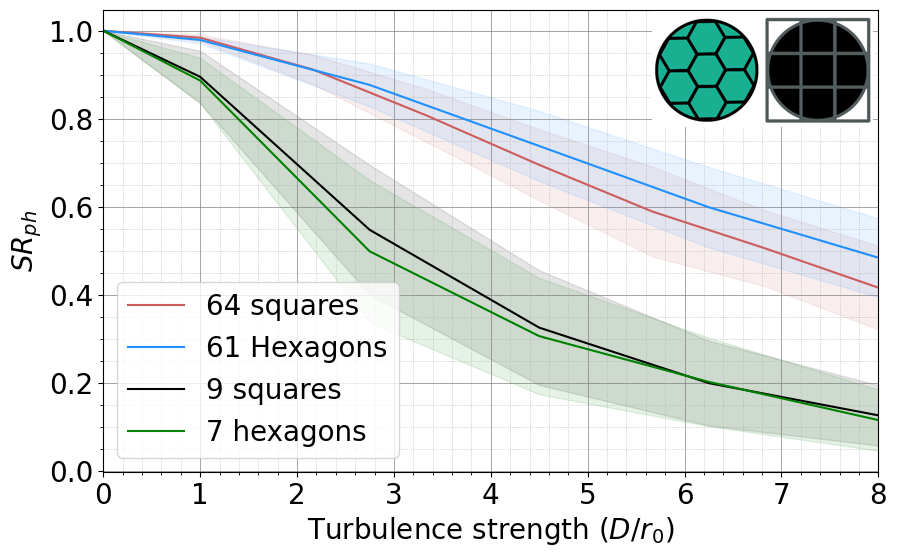}
        \label{subfig: SRvsDr0}
     \end{subfigure}
     \begin{subfigure}[b]{0.49\textwidth}
        \centering
        \includegraphics[width=\linewidth]{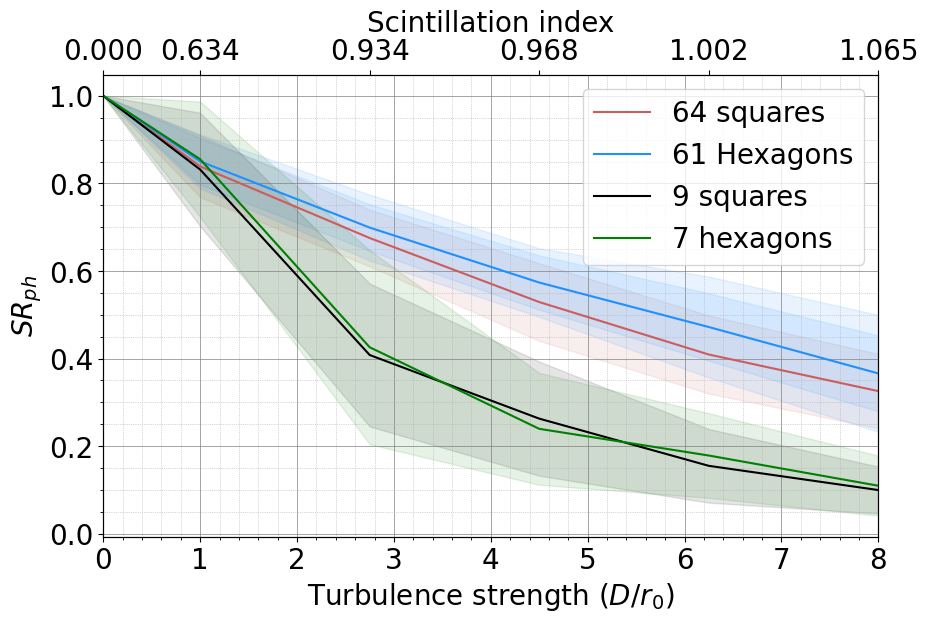}
        \label{subfig: SRvsDr0_scinti}
     \end{subfigure}
    \caption{Strehl ratio of the hexagonal and square arrays as a function of the turbulence strength, $D/r_0$. Left: devices performing free of scintillation. Right: devices performing under scintillation. The top abscissa indicates the scintillation index.}
    \label{fig: SRvsDr0_results}
\end{figure}

\begin{figure}[hbt!]
    \centering
    \includegraphics[width=0.75\linewidth]{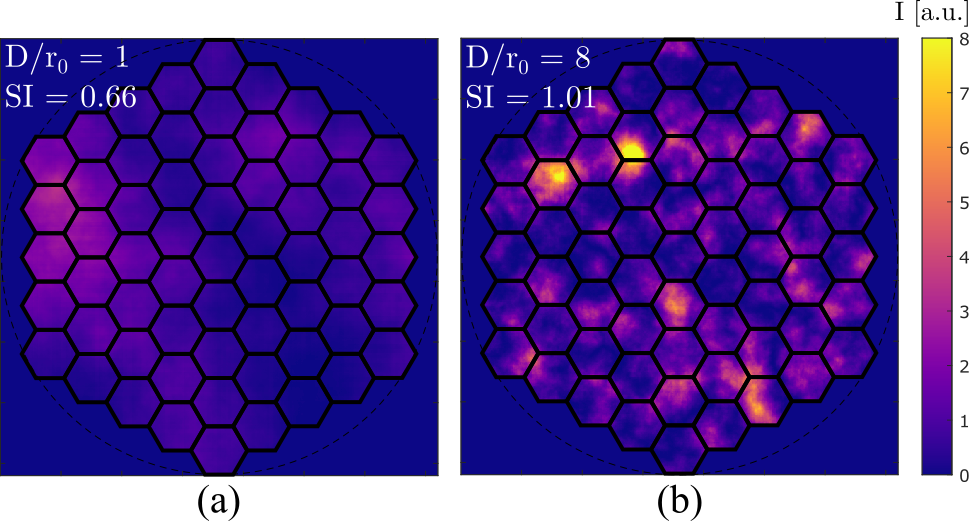}
    \caption{Distribution of the optical intensity at the telescope pupil with overlaid subapertures for a $61$ hexagons array. The wavefront that produced the distribution in (a) has $D/r_0 = 1$, while for (b) $D/r_0 = 8$.}
    \label{fig:pupilIntensityMap}
\end{figure}

The simulation results for the degradation of the $\mathit{SR}_{\mathit{ph}}$, as the atmospheric turbulence gets stronger, are shown in Fig. \ref{fig: SRvsDr0_results}. 
The comparison is made between the $3\times 3$ subapertures square array and the $7$ subapertures ($1$-ring) hexagonal array, and then between the $8\times 8$ subapertures square array and the $61$ subapertures ($4$-rings) hexagonal array. 
The results shown in Fig. \ref{fig: SRvsDr0_results}b are for when scintillation effects are included in the pipeline where the scintillation index, $\mathit{SI}$, indicated on the top abscissa for each $D/r_0$ value, measures the normalized variance in the intensity $I$:

\begin{equation}
\mathit{SI} = \frac{\langle I^2 \rangle - \langle I \rangle^2}{\langle I \rangle^2}.    
\end{equation}

\begin{figure}[hbt!]
    \centering
    \includegraphics[width = \textwidth]{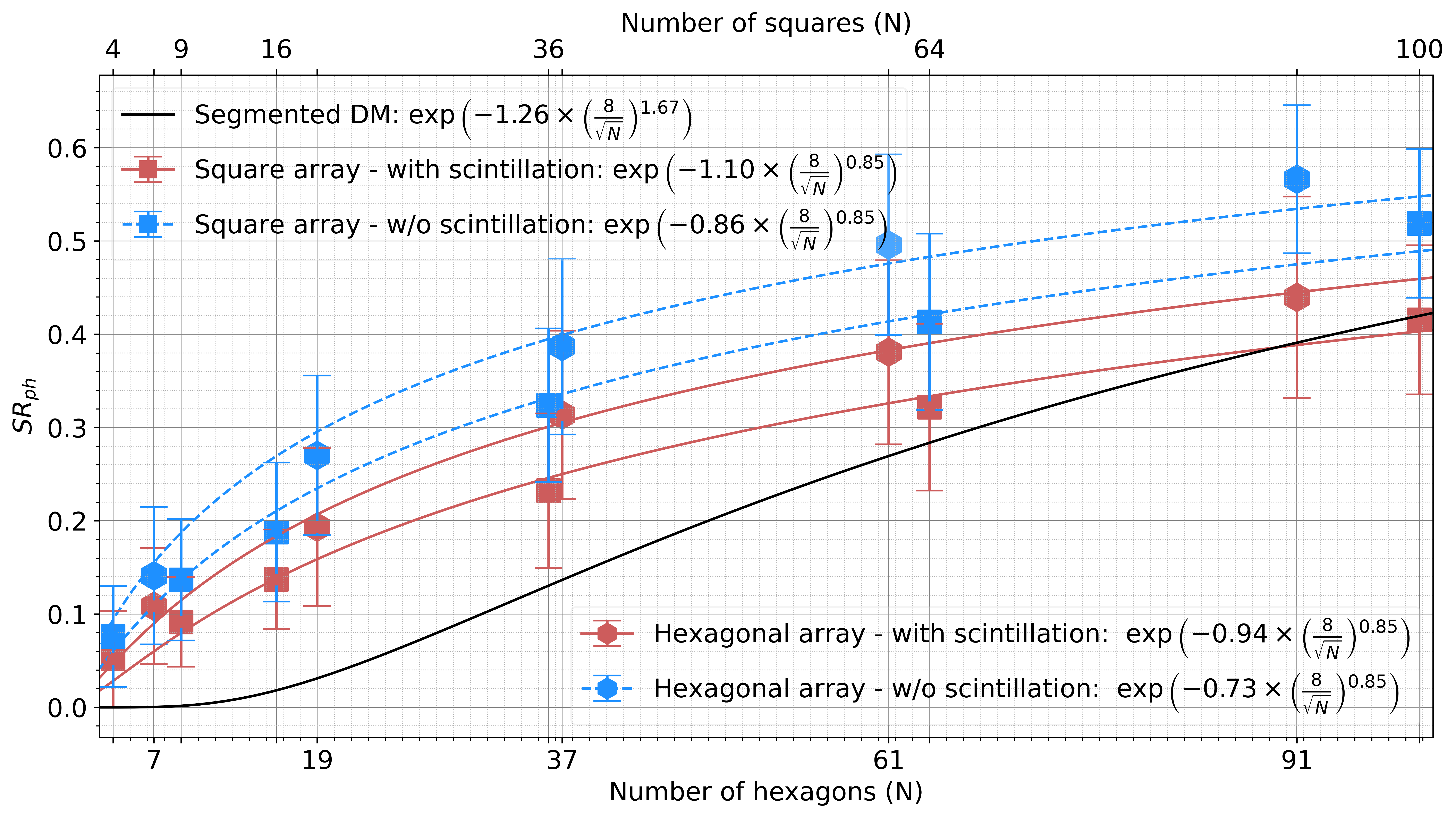}
    \caption{Strehl ratio of hexagonal and square arrays as a function of subapertures at $D/r_0 = 8$. The aperture size is kept constant at $D = 0.4~$m for both arrangements. The bottom abscissa is for the total number of subapertures in hexagonal arrays while the top abscissa shows the same for square arrays.}
    \label{subfig: SRvsArr}
\end{figure}

The worst-case scintillation at $D/r_0 = 8$ results from propagating the phase screens $50~$km in free space along the line-of-sight (see Fig. \ref{fig:pupilIntensityMap}). This is the propagation distance during the lowest elevation stage of a typical LEO-to-ground downlink. At the diffraction limit ($D/r_0 = 0$), the $\mathit{SR}_{\mathit{ph}}$ is unity by definition, and $\mathit{SI} = 0$. 
The drop in $\mathit{SR}_{\mathit{ph}}$ is steeper for the cases with fewer subapertures as expected.
From Fig. \ref{subfig: SRvsArr} it can be seen that the hexagonal array always performs better under strong turbulence conditions while using fewer subapertures. 
This advantage is due to the higher fill factor that hexagonal arrays have on the telescope's circular aperture.   

An essential quality of wavefront correctors is their ability to fit and hence correct the aberrated wavefront. The fitting error of DMs was calculated by Hudgin \cite{hudgin_wave-front_1977}. 
For segmented DMs, it depends on the number of actuators and the arrangement of the segments.
The mean-square fitting error is 

\begin{equation}
\label{eq: we_error}
\sigma^2 = \alpha \left(\frac{D}{Mr_0}\right)^{5/3},    
\end{equation}
where $M$ is the number of segments along one dimension and $\alpha$ is a constant that depends on the influence function. For piston-only actuators $\alpha = 1.26$ rad$^2$.
The $\mathit{SR}$ falls off exponentially with $\sigma^2$ according to Mar\'echal approximation ($\mathit{SR} = e^{-\sigma^2}$), providing a way to directly compare the photonic corrector to segmented DMs.  
The data points in Fig. \ref{subfig: SRvsArr} have $D/r_0 = 8$.
Fitting an exponential function with one independent variable, $M$, and three constant parameters, we write $\mathit{SR}_{\mathit{ph}}$ as 
\begin{equation}
\label{eq: SR_we_error}
\mathit{SR}_{\mathit{ph}} = \exp\left(-\sigma^2\right) = \exp\left[- s \cdot \alpha \left( \frac{8}{M} \right)^{\beta}\right].
\end{equation}

We find out that our photonic wavefront corrector with a square array follows an adjusted law with $\alpha = 0.86$ rad$^2$ and $\beta = 0.85$. The parameter $s$, which we name the \textit{scintillation coefficient}, accounts for scintillation effects. It is calculated by finding the multiplication factor between the with and without scintillation curves in Fig. \ref{subfig: SRvsArr}.  Its value is $1.28$ when the corrector is operating under $D/r_0 = 8$ ($\mathit{SI} \approx 1$). 
Otherwise, its value is unity for negligible scintillation. 
Table \ref{tab: alpha_vals} lists the values for $\alpha$ and $s$ for both configurations. A comparison between a toy model of our concept and circular DMs that perfectly remove the low-order Zernike modes as estimated by Noll \cite{noll_zernike_1976} was given in \cite{diab2022photonic}. 


\begin{table}[hbt!]
    \centering
    \resizebox{\columnwidth}{!}{
    \begin{tabular}{c|c|c|c}
     & Square array PIC & Hexagonal array PIC & Segmented DM \\ \hline 
     Scintillation coefficient, $s$ & 1.28 & 1.28 & 1\\ \hline
     Fitting error coefficient, $\alpha$ [rad$^2$]  & 0.86 & 0.73 & 1.26 \\ \hline
     Exponent, $\beta$ & 0.85 & 0.85 & 1.67\\ 
    \end{tabular}
    }
    \caption{Values of $s$, $\alpha$, and $\beta$ for the fitting error in Eq. \ref{eq: SR_we_error}.
}
    \label{tab: alpha_vals}
\end{table}

Notice that $\mathit{SR}_{\mathit{ph}}$ only quantifies the correction quality and is not sensitive to the coupling and propagation losses in the PIC. 
To characterize the total throughput of the corrector, the $\mathit{SR}_{\mathit{ph}}$ must be multiplied by the total throughput of the PIC to estimate the output optical power provided by the system. 
The expected total throughput of the device simulated in this paper is $\sim 0.1$, but optimized grating couplers with an $> 80\%$ efficiency \cite{roelkens_high_2007} and an SWG mode expander with a $> 92\%$ efficiency \cite{cheben_broadband_2015} could boost the total throughput up to $> 0.5$ at the cost of a more complex fabrication process. The propagation loss can be reduced by switching to a silicon nitride (SiN) platform.
We chose to simulate sub-optimum devices to predict the performance of our first-generation PIC that we will use to prove the concept of photonic AO correction experimentally.

\section{Conclusions and future work}
\label{sec: conclusion}
We performed simulations to predict the total optical power delivered by silicon photonics wavefront correctors designed to efficiently couple light into SMFs in the presence of atmospheric turbulence. 
The simulations explored the parameter space of the design geometries and produced estimates for the quality of correction expected from devices with different sizes and under the various turbulence conditions expected in LEO-to-ground FSO links. 

In terms of fitting errors, PICs with hexagonal arrays operating in the presence of scintillation effects are expected to provide an $\mathit{SR}_{\mathit{ph}}$ that is $\sim 1.4$ times higher than that provided by segmented DMs.
Moreover, the temporal errors of the photonic concept are much smaller thanks to the fast rise time of the phase shifters. 
Continuous facesheet mirrors perform better than both approaches; however, they are limited in stroke, along with lower control bandwidths and higher inter-actuator crosstalk \cite{ravensbergen_deformable_2009, stroebele_deformable_2016}.
PICs are not immune to crosstalk, but the thermal crosstalk between the TOPSs in the PIC can be mitigated by adding trenches and suspended waveguides to thermally isolate them. Electro- and piezo-optic modulators could also be used to eliminate crosstalk, which additionally facilitates the placement of photonic wavefront correctors inside cryostats for infrared astronomical instrumentation. 


As in a classical AO system, the photonic corrector requires a WFS and a controller to measure and reconstruct the wavefront before the commands for co-phasing the beams can be calculated. 
Two wavefront sensing schemes are proposed: an open-loop Shack-Hartmann WFS that samples and senses the incoming wavefront before coupling into the PIC, and an integrated binary tree of Mach-Zehnder interferometers (MZIs) \cite{milanizadeh_coherent_2021} to measure and directly correct the relative phase errors on-chip. 
The results from simulations and experimental work that includes the WFS in the pipeline will be reported in a future publication.

Apart from the simulation work reported here, we ran lab experiments on smaller $2\times 2$ square arrays using multiple phase plates to introduce atmospheric distortions in the beam and relay optics to image the wavefront on the MLA. The loop was closed on these arrays using a sensorless gradient descent algorithm that maximizes the power in the output SMF.
We also designed and fabricated larger arrays that we plan to test in combination with an external SH-WFS and integrated MZI trees. 
The results from the experimental work will also be reported in future communications. 
Later, field-testing on ground-to-ground analog links should help raise the system's technological readiness level (TRL) and prepare the setup for establishing links with LEO satellites equipped with FSO laser terminals.  

\begin{backmatter}

\bmsection{Funding}
National Research Council Canada (HTSN 628, HTSN 647).

\bmsection{Acknowledgments}
We would like to acknowledge CMC Microsystems for the provision of products and services that facilitated this research. S.S. acknowledges the support of the Natural Sciences and Engineering Research Council of Canada (NSERC) discovery grant.

\bmsection{Disclosures} 
The authors declare no conflicts of interest.

\bmsection{Data availability} 
Data underlying the results presented in this paper are not publicly available at this time but may be obtained from the authors upon reasonable request.

\end{backmatter}

\bibliography{paper}

\end{document}